\documentclass[letter]{IEEEtran}
\usepackage{amsmath,amssymb,amsfonts}
\usepackage{algorithmic}
\usepackage{graphicx}
\usepackage{textcomp}
\def\BibTeX{{\rm B\kern-.05em{\sc i\kern-.025em b}\kern-.08em
    T\kern-.1667em\lower.7ex\hbox{E}\kern-.125emX}}
\usepackage{caption}
\usepackage{tabularx}
\usepackage{float}
\usepackage{cite}

\usepackage{mathtools}
\usepackage[hidelinks]{hyperref}
\usepackage{tikz}
\usetikzlibrary{arrows.meta, decorations.pathmorphing, positioning,quotes}

\usepackage{cleveref}
\usepackage{bm}
\usepackage{wasysym}
\usepackage{amsthm}

\theoremstyle{plain}
\newtheorem{theorem}{Theorem}
\newtheorem{proposition}{Proposition}
\newtheorem{lemma}{Lemma}

\theoremstyle{definition}
\newtheorem{definition}{Definition}

\newtheorem{assumption}{Assumption}
\newtheorem{notation}{Notation}
\newtheorem{remark}{Remark}

\Crefname{assumption}{Assumption}{Assumptions}

\newcommand{\X}{\mathcal X}
\newcommand{\Set}{\mathcal S}
\newcommand{\rn}{\mathbb R^n}
\newcommand{\rnn}{\mathbb R^{n\times n}}
\newcommand{\rb}{\mathbb R}

\newcommand{\Dn}{\mathcal D_n}
\newcommand{\F}{\mathcal F}
\newcommand{\An}{\mathcal A_{n}}
\newcommand{\A}{\mathcal A}
\newcommand{\T}{^\top}
\newcommand{\Pn}{\mathcal P_n}
\newcommand{\Cn}{\mathcal C_n}

\mathtoolsset{showonlyrefs}

\begin{document}
\title{Linear Lyapunov Functions for Nonlinear Compartmental Systems}
\author{Sondre Wiersdalen, Mike Pereira, Annika Lang, G\'abor Szederk\'enyi, Jean Auriol, and Bal\'azs Kulcs\'ar.
  \thanks{Received 31 July 2025; revised 1 March 2026; accepted 12 June 2026. Date of publication dd Month yyyy; date of current version dd Month yyyy. This work has partially been supported by the Transport Area of Advance, at Chalmers University of Technology, Gothenburg, Sweden, by the project Learning in Stochastic Traffic Networks, Chalmers University of Technology, by the Swedish Research Council (VR) through grant no.\ 2020-04170, by the Wallenberg AI, Autonomous Systems and Software Program (WASP) funded by the Knut and Alice Wallenberg Foundation, by the Chalmers AI Research Centre (CHAIR), and by the European Union (ERC, StochMan, 101088589). (Corresponding Author: Sondre Wiersdalen)}
  \thanks{Sondre Wiersdalen is with the Department of Electrical Engineering, Chalmers University of Technology, SE-412 96 Gothenburg, Sweden (e-mail: chanon@chalmers.se).}
  \thanks{Mike Pereira is with the Department of Geosciences and Geoengineering, Mines Paris -- PSL University, 75272 Paris, France (e-mail: mike.pereira@minesparis.psl.eu)}
  \thanks{Annika Lang is with the Department of Mathematical Sciences, Chalmers University of Technology \& University of Gothenburg, SE-412 96 Gothenburg, Sweden (e-mail: annika.lang@chalmers.se).}
  \thanks{G\'abor Szederk\'enyi is with the Faculty of Information Technology and Bionics, P\'azm\'any P\'eter Catholic University, Szentkir\'alyi u. 28, 1088 Budapest, Hungary (e-mail: szederkenyi@itk.ppke.hu).}
  \thanks{Jean Auriol is with Universit\'e Paris-Saclay, CNRS, CentraleSup\'elec, Laboratoire des Signaux et Syst\`emes, 91190, Gif-sur-Yvette, France (e-mail: jean.auriol@centralesupelec.fr).}
  \thanks{Bal\'azs Kulcs\'ar is with the Department of Electrical Engineering, Chalmers University of Technology, SE-412 96 Gothenburg, Sweden (e-mail: kulcsar@chalmers.se).}
}
\maketitle
\begin{abstract}
  This technical note examines exponential stability of the null solution to a large class of compartmental systems governed by ordinary differential equations.
  Sufficient conditions under which these systems admit a linear  Lyapunov function are provided.
  The coefficients of the Lyapunov functions and the exponential decay rate they yield are obtained from an eigenvalue problem.
  For a special case of the system class considered, we derive an equivalence between attractivity of the null solution and the existence of a linear Lyapunov function.
\end{abstract}

\section{Introduction}
An abundance of dynamical systems are governed by mass balance equations.
Examples of such systems include (but are not limited to) macroscopic traffic flow models \cite{daganzo94,liptak2021,pereira2022parameter,pereira2022short}, epidemiological models \cite{muldowney95,yorke1976}, biological systems describing blood glucose \cite{bergman89} or the flow of ribosomes \cite{margaliot2012}, and even probabilistic models like Markov chains \cite[ch. 12]{walter1999}.
The goal of this paper is to investigate exponential stability of the null solution to compartmental systems governed by ordinary differential equations.

Compartmental systems are, in essence, mass-balance equations and describe the flow of some nonnegative quantity between so-called compartments \cite{jacquez93}.
A \emph{compartment} can be viewed as a placeholder for the nonnegative quantity (such as mass) that traverses between them.
These systems are in general nonlinear and nonautonomous, and several authors have studied their asymptotic behavior.
Asymptotic stability of the origin and washout are considered in \cite{eisenfeld1982washout}.
Sufficient criteria for the existence, uniqueness, and asymptotic stability of equilibrium points and periodic solutions were proposed in \cite{sandberg1978}.
Under mild conditions, \cite{maeda1978asymptotic} showed that autonomous compartmental systems do not admit periodic solutions.
Quite general criteria for asymptotic stability of the origin for linear time-varying compartmental systems are given in \cite{aronsson1978differential}.

Few works have considered the convergence rate of these asymptotic behaviors.
An exception is \cite[Theorem 4.1]{eisenfeld1982washout}, which establishes exponential stability for autonomous compartmental systems.
Even less attention has been given to stability for compartmental systems that are both nonautonomous \emph{and} nonlinear.
One notable example is \cite[Theorem 1]{ladde1976a}, which considers connective stability.

The main contributions and layout of this paper are as follows.
In \Cref{sec:compartmental} we introduce nonlinear nonautonomous compartmental systems, as defined in \cite{jacquez93}, for which exponential stability of the null solution is considered in \Cref{sec:main}.
This stability property is established by means of linear Lyapunov functions.
In \Cref{sec:matrix}, we provide sufficient conditions for the existence of such functions.
A converse Lyapunov theorem is provided in \Cref{sec:converse} for a special case of compartmental systems.
The theoretical results are demonstrated and discussed in \Cref{sec:sim}, followed by the conclusion in \Cref{sec:conclusion}.
All proofs are given in the Appendix.

\textit{Notation:}
Let $\mathbb N\coloneqq\{1,2,\dots\}$ and $\rn$ denote Euclidean space.
We write $x\coloneqq(x_1,\dots,x_n)$ to denote vectors in $\rn$.
The vectors $(1,\dots,1)$ and $(0,\dots,0)$ in $\rn$ are denoted by $\bm 1$ and $\bm 0$, respectively, and the dimension is taken from context.
For any $x,y\in \mathbb R^n$ we write $x \leq y$ if $y-x\in[0,+\infty)^n$ and we write $x\geq y$ if $x-y\in[0,+\infty)^n$.
If $x\in \rn$, then $x$ is interpreted as a column vector unless stated otherwise.
If $x$ is a real-valued vector or matrix, then its transpose is denoted by $x \T$.
If $f$ is a matrix (or a matrix-valued function), then $f_{ij}$ denotes the element in the $i$-th row and $j$-th column of $f$.
The set of all permutation matrices in $\rnn$ is denoted by $\Pn$.

\section{Compartmental Systems}\label{sec:compartmental}
Let us recall compartmental systems as described in \cite{jacquez93}.
Consider a system of $n>1$ first-order ordinary differential equations:
\begin{equation}
 \label{eq:C}\tag{C}
  \dot x(t) = f(t,x(t))x(t)
\end{equation}
where $f=(f_{ij})$ is an $n\times n$ matrix-valued function and $x(t)\in \rn$.
In \cite{jacquez93}, system~\eqref{eq:C} is known as a \emph{compartmental system} if $f$ is continuous and its elements $f_{ij}$ satisfy
\begin{equation}\label{eq:jac}
  f_{ij}(t,q)\geq 0,\; i\neq j\quad \sum_{i=1}^nf_{ij}(t,q)\leq 0,\; j=1,\dots,n,
\end{equation}
for all $t\geq 0$ and $q\in \rn$ such that $f(t,q)$ is defined.
By introducing the functions $f_{0j}\coloneqq -\sum_{i=1}^nf_{ij}$, $j=1,\dots,n$, we rewrite \eqref{eq:C} as $n$ mass-balance equations
\begin{equation}
  \label{eq:balance}\tag{C$'$}
  \dot x_i=-f_{0i}x_i-\sum_{j\neq i}f_{ji}x_i+\sum_{j\neq i}f_{ij}x_j,\quad i=1,\dots,n,
\end{equation}
where arguments are omitted and $x_i$ denotes the mass (or other nonnegative quantity) in the $i$-th compartment over time.

Let ``compartment $0$'' denote the so-called \emph{environment}.
Then $f_{ij}x_j$ denotes the flow from compartment $j\neq 0$ to compartment $i\neq j$ over time.\footnote{Since there are no flows from the environment to any of the compartments, \eqref{eq:C} is also known as an inflow-closed compartmental system \cite[p. 45]{jacquez93}.}
To illustrate \eqref{eq:C}, we can draw a \emph{connection diagram}, as shown in \Cref{fig:diagram}.
Each dot labeled with $i$ represents compartment $i$.
If $f_{ij}\not \equiv 0$, $i\neq 0$, and $i\neq j$, then an arrow is drawn from $j$ to $i$.
If $f_{0j}\not \equiv 0$, then an arrow originates from $j$ but is not connected to any other compartment.
\begin{figure}[H]
  \centering
  \begin{tikzpicture}[
    node distance = 10mm and 20mm,
    compartments/.style = {circle, fill = black, inner sep = 1.5pt, label distance = 1mm},
    environment/.style = {circle, fill = white},
    every edge quotes/.style = {font = \footnotesize, sloped}
    ]
    \begin{scope}[nodes=compartments]
    \node[label=above:{$1$}] (1)  {};
    \node[label=above:{$2$}] (2) [right=of 1] {};
    \node[label=above:{$3$}] (3) [right=of 2] {};
    \node[label=above:{$4$}] (4) [right=of 3] {};
    \end{scope}
    \draw (1) edge["$f_{21}$",->,bend left, auto]  (2);
    \draw (2) edge["$f_{12}$",->,bend left,auto = left]  (1);
    \draw (2) edge["$f_{32}$",->,bend left, auto]         (3);
    \draw (3) edge["$f_{23}$",->,bend left,auto = left]   (2);
    \draw (3) edge["$f_{43}$",->,bend left, auto]         (4);
    \draw (4) edge["$f_{34}$",->,bend left,auto = left]   (3);
    \begin{scope}[nodes=environment]
      \node (01) [below = of 1] {};
      \node (04) [below = of 4] {};
    \end{scope}
    \draw (1) edge["$f_{01}$",->, auto = right]  (01);
    \draw (4) edge["$f_{04}$",->, auto = left]  (04);

  \end{tikzpicture}
  \caption[mycaption]{
    An example of a connection diagram for \eqref{eq:C}, indicating that $f(t,q)$ is tridiagonal for all $t$ and $q$.
     }\label{fig:diagram}
   \end{figure}
The following assumption holds throughout the rest of the article.
\begin{assumption}\label{A1}
  Given $\X\subset [0,+\infty)^n$, $\tilde D\subset \rb^{n+1}$, and $f:\tilde D\to\rnn$, it is assumed that
  \begin{enumerate}
  \item  $D\coloneqq [0,+\infty)\times \X\subset \tilde D$ and $[0,\epsilon)^n\subset \X$, for some $\epsilon>0$,
  \item $f$ is continuous and satisfies \eqref{eq:jac}  for all $(t,q)\in D$,
  \item for all $\xi\in \X$, \eqref{eq:C} admits a unique solution\footnote{By solution we mean a function $x:[0,+\infty)\to \rn$ that satisfies \eqref{eq:C} for all $t>0$.} $x$ such that~$x(0)=\xi$ and $x(t)\in \X$ for all $t\geq 0$.\footnote{Granted 1) and 2), assumption 3) is not needed explicitly if $\tilde D=\rb^{n+1}$, $\X=[0,+\infty)^n$, and $(t,q)\mapsto f(t,q)q$ is Lipschitz in $q$ uniformly in $t$ \cite{wyatt1985}.}
  \end{enumerate}
\end{assumption}
When convenient, we use $t\mapsto \phi(t,\xi)$ to denote the solution to \eqref{eq:C} with $\phi(0,\xi)=\xi$.
Clearly, the \emph{null solution} $x\equiv (0,\dots,0)$ satisfies \eqref{eq:C}.
The definitions for stability we use are as follows.
\begin{definition}[\cite{khalil2002nonlinear}]\label{def:stability}
  Let $\Set \subset \X$ be nonempty.
  The null solution to~\eqref{eq:C} is said to be
  \begin{itemize}
  \item \emph{attractive with respect to} $\Set$ if $\lim_{t\to+\infty}|\phi(t,\xi)|_1=0$ for all $\xi\in \Set$,
  \item \emph{exponentially stable} (ES) \emph{with respect to} $\Set$ if there exist positive real numbers $\lambda$ and $\gamma$  such that
    \begin{equation}\label{eq:ES}
    |\phi(t,\xi)|_1\leq \gamma e^{-\lambda t}|\xi|_1 \text{ for all } t\geq 0 \text{ and } \xi\in \Set.
  \end{equation}
  \end{itemize}
\end{definition}
We refer to any $\lambda>0$ satisfying \eqref{eq:ES} as an \emph{exponential decay rate}, and we seek such rates for the system \eqref{eq:C}.
The main tool we use to assess the stability for \eqref{eq:C} is the following.
\begin{proposition}\label{prop:lyap}
  If there exist $\lambda>0$ and $v\in (0,+\infty)^n$ such that
  \begin{equation}
  \label{eq:lyap}
  v\T f(t,q)\leq -\lambda v\T,\quad (t,q)\in D,
  \end{equation}
  then the null solution to \eqref{eq:C} is ES with respect to $\X$ and \eqref{eq:ES} is satisfied with $\mathcal S \coloneqq \X$ and $\gamma \coloneqq \frac{\max_i v_i}{\min_i v_i}$.
\end{proposition}
\begin{remark}\label{rem:lyap}
  If the hypotheses of \Cref{prop:lyap} are true, then~$q\mapsto v\T q$ is a (linear) Lyapunov function of \eqref{eq:C} on $\X$ in the sense of LaSalle \cite{lasalle1968}.
  If $f\equiv F$ for some $F\in \rnn$ such that \Cref{A1} holds, then hypotheses of \Cref{prop:lyap} are true if and only if\footnote{A formula for $\lambda$ and $v$ is derived in the Appendix} $F$ is nonsingular \cite{rantzer2015}, \cite{jacquez93}.
\end{remark}
In the sequel, we seek suitable conditions to verify exponential stability of the null solution to \eqref{eq:C} by means of \Cref{prop:lyap}.
The hypotheses in \Cref{prop:lyap} are sufficient to assert exponential stability, but not necessary.
Under additional assumptions on $f$, however, we show in \Cref{sec:converse} that \eqref{eq:lyap} is equivalent to the null solution being attractive with respect to $\X$.

\section{Main Results}\label{sec:main}

\subsection{Analysis of Compartmental Matrices}\label{sec:matrix}
A matrix $F\in \rnn$ is said to be \emph{compartmental} \cite{jacquez93}, \cite{deLeenheer2001}, if
\begin{equation}
  \label{eq:Cn}
  F_{ij}\geq 0,\quad i\neq j,\quad \sum_{i=1}^nF_{ij}\leq 0,\quad j=1,\dots,n.
\end{equation}
Let $\Cn$ denote the set of compartmental matrices in $\rnn$.
By \Cref{A1}, $f$ maps $D$ into $\Cn$.
Hence we pose the following problem given a set $\F\subset \Cn$:
\begin{equation}\label{eq:p1}\tag{P}
  \begin{cases}
    \text{Find  $\lambda>0$ and $v\in(0,+\infty)^n$ such}\\
    \text{that $v\T F\leq -\lambda v\T $ for all $F\in \F$.}
  \end{cases}
\end{equation}
\begin{remark}
  If \eqref{eq:p1} admits a solution for $\F\coloneqq \{f(t,q)\;:\; (t,q)\in D\}$, it follows from \Cref{prop:lyap} that the null solution to \eqref{eq:C} is ES with respect to $\X$.
\end{remark}
To address \eqref{eq:p1}, we introduce a special class of compartmental matrices.
\begin{definition}\label{def:ds}
  Let $\Dn$ denote the set of all $F\in \Cn$ such that $\sum_{i=1}^nF_{in}<0$ and for all $1\leq j<n$,
  \begin{equation}
    \label{eq:down}
      \sum_{i=1}^nF_{ij}<0 \text{ or } F_{ij}>0 \text{ for some } i>j.
    \end{equation}
If $F\in\Dn$, then $F$ is said to be \emph{downstream connected}.
\end{definition}
\begin{notation}
  For convenience, define the maps $d^n:\Cn\to \rn$, $u^n_j:\Cn\to \rb$ by
\begin{align}
  \label{eq:dj}
      d^n_j(F)&\coloneqq
  \begin{cases}
  -\sum_{i=1}^nF_{ij}+\sum_{i=j+1}^nF_{ij},\quad &j<n,\\
-\sum_{i=1}^{n}F_{in},\quad &j=n,
  \end{cases}\\
  u_j^n(F)&\coloneqq \text{$\sum_{i=1}^{j-1}F_{ij}$},\quad j=2,\dots,n. \label{eq:uj}
\end{align}
Let $F\in\rnn$ be compartmental.
Then it follows by definition of $d^n$ and $\Dn$ that $d^n(F)\in(0,+\infty)^n$ if and only if $F\in \Dn$.
\end{notation}

A compartmental matrix can be determined to be downstream connected by inspecting its directed graph.
Define the directed graph $\Gamma(F)$ of $F\in \Cn$  by the nodes $\{1,\dots,n\}$ and the arcs $\{(j,i)\;:\; F_{ij}\neq 0\}$.
Node $j$ is an \emph{outflow node} if the $j$-th column sum of $F$ is negative, denoted by a double circle in \Cref{fig:graph}.

\begin{figure}[ht!]
  \begin{center}
   \begin{tikzpicture}
    \node[style=circle, draw=black, minimum size = 0.7cm]         (1) at (0,0) {$1$};
    \node[style=circle, draw=black, minimum size = 0.7cm]         (2) at (1.5,0) {$2$};
    \node[style=circle, draw=black, minimum size = 0.7cm]         (3) at (3,0) {$3$};
    \node[style=circle, draw=black, minimum size = 0.7cm]         (i) at (4.5,0) {$i$};
    \node[style=circle, draw=black, minimum size = 0.7cm, double]  (n) at (6,0) {$n$};
    \draw (1) edge[->] (2);
    \draw (2) edge[->] (3);
    \draw (3) edge[->, dashed] (i);
    \draw (i) edge[->, dashed] (n);

    \draw (1) edge[<-,bend left, below] (2);
    \draw (1) edge[<-,bend left, below] (3);
    \draw (1) edge[<-,bend left, below] (i);
    \draw (1) edge[<-,bend left, below] (n);
    \node[] (ghost) at (0,-1/2) {};
  \end{tikzpicture}

  \begin{tikzpicture}
    \node[style=circle, draw=black, minimum size = 0.7cm]  (1) at (0,0) {$1$};
    \node[style=circle, draw=black, minimum size = 0.7cm]         (2) at (1.5,0) {$2$};
    \node[style=circle, draw=black, minimum size = 0.7cm]         (3) at (3,0) {$3$};
    \node[style=circle, draw=black, minimum size = 0.7cm]         (i) at (4.5,0) {$i$};
    \node[style=circle, draw=black, minimum size = 0.7cm, double]         (n) at (6,0) {$n$};
    \draw (1) edge[<-,bend left,  above] (2);
    \draw (1) edge[->,bend right, above] (2);
    \draw (2) edge[<-,bend left,  above] (3);
    \draw (2) edge[->,bend right, above] (3);
    \draw (3) edge[<-,bend left,  above, dashed] (i);
    \draw (3) edge[->,bend right, above, dashed] (i);
    \draw (i) edge[<-,bend left,  above, dashed] (n);
    \draw (i) edge[->,bend right, above, dashed] (n);
    \node[] (ghost) at (0,-1/2) {};
  \end{tikzpicture}
   \begin{tikzpicture}
    \node[style=circle, draw=black, minimum size = 0.7cm]         (1) at (0,0) {$1$};
    \node[style=circle, draw=black, minimum size = 0.7cm]         (2) at (1.5,0) {$2$};
    \node[style=circle, draw=black, minimum size = 0.7cm]         (3) at (3,0) {$3$};
    \node[style=circle, draw=black, minimum size = 0.7cm]         (i) at (4.5,0) {$i$};
    \node[style=circle, draw=black, minimum size = 0.7cm, double]  (n) at (6,0) {$n$};
    \draw (1) edge[<-] (2);
    \draw (2) edge[<-] (3);
    \draw (3) edge[<-, dashed] (i);
    \draw (i) edge[<-, dashed] (n);

    \draw (1) edge[->,bend right, below] (n);
    \draw (2) edge[->,bend right, below] (n);
    \draw (3) edge[->,bend right, below] (n);
    \draw (i) edge[->,bend right, below] (n);
    \node[] (ghost) at (0,-1/2) {};
  \end{tikzpicture}
  \end{center}
  \caption[mycaption]{
    Directed graphs associated with some matrices in $\Cn$.
    All are downstream connected because $n$ is an outflow node and every node $j<n$ has an arc to some $i>j$.

     }\label{fig:graph}
   \end{figure}
   \Cref{def:ds} is motivated by the following proposition.
 \begin{proposition}\label{prop:ds}
   Let $\F\subset \Cn$ be nonempty, let $P\in \Pn$, and suppose $(\lambda,v)$ solves \eqref{eq:p1}.
   If $(P\T v)_1\geq \dots \geq (P\T v)_n$, then
   \begin{equation}\label{eq:F-permute}
     d^n(P\T FP)\geq(\lambda,\dots,\lambda),\quad F\in \F
   \end{equation}
   and $P \F P\T$ is a set of downstream connected matrices.
 \end{proposition}
 \Cref{prop:ds} shows that whenever \eqref{eq:p1} admits a solution for a set $\F\subset \Cn$, then there is a \emph{single} permutation matrix such that $P F P\T$ is downstream connected for every $F\in \F$.

 Let $n>1$ be an integer, let $a,b>0$ be real, and define the $(n\times n)$ downstream connected matrix\footnote{The directed graph of $g_n(a,b)$ is depicted (first from above) in \Cref{fig:graph}.}
    \begin{equation}\label{eq:gn}
g_n(a,b)\coloneqq \begin{bmatrix}
 -a & b   & b   & \cdots & b \\
\hphantom{-}a      & -a-b &  0   & \cdots &  0 \\
\hphantom{-}  0      & a   & -a-b & \ddots &  \vdots \\
\hphantom{-}\vdots &  \ddots & \ddots & \ddots & 0 \\
\hphantom{-} 0  & \cdots & 0 &a & -a-b
\end{bmatrix}.
\end{equation}
It is shown in \Cref{appendix:A} that the (constrained) eigenvalue problem
\begin{equation}\label{eq:evp} \tag{EVP}
  \begin{cases}
    \text{find $\lambda>0$ and $v\in(0,+\infty)^n$ such}\\
    \text{that $v\T g_n(a,b)=-\lambda v\T $ and $v_1=1$}\\
  \end{cases}
\end{equation}
admits a unique solution.\footnote{In fact, $v$ is a left Perron-Frobenius eigenvector associated with $-\lambda$.}
Associate with $g_n(a,b)$ the set
\begin{equation}\label{eq:An}
  \begin{split}
    \An(a,b)\coloneqq \big \{&A\in \Cn :v\T A\leq -\lambda v\T  \text{ where }(\lambda,v)\\
    &\text{ is the unique solution to \eqref{eq:evp}}\big \}.
  \end{split}
\end{equation}
\begin{remark}
  It follows from definition \eqref{eq:An} of $\An(a,b)$ that the solution to \eqref{eq:evp} is a solution to \eqref{eq:p1} if $a,b>0$ and $\F= \An(a,b)$.
\end{remark}
In the following theorem we provide a simple test to verify that a compartmental matrix belongs to $\An(a,b)$ and we provide a necessary and sufficient test for the existence of a solution to \eqref{eq:p1} when $\F\subset \Cn$ is bounded.
Note that the set $\An(a,b)$ itself is not bounded.
\begin{theorem}\label{thm:matrix}
 i) If $a,b>0$, $F\in \Cn$, and
    \begin{equation}\label{eq:djuj}
     d^n_j(F)\geq a,\;j\geq 1,\qquad  u_j^n(F)\leq b,\; j>1,
   \end{equation}
   then $F\in \An(a,b)$.
   
ii) Problem \eqref{eq:p1} has a solution for bounded $\F\subset \Cn$ if and only if there exist $a,b>0$ and $P\in \Pn$ such that
\begin{equation}\label{eq:F-include}
 P \T \F P\subset \An(a,b).
\end{equation}
\end{theorem}
\begin{remark}\label{rem:matrix}
  Consider system \eqref{eq:C} with \Cref{A1}.
  Fix any $P\in \Pn$ and $\epsilon>0$ and suppose
  \begin{align}\label{eq:aj}
    a\coloneqq &\inf_{(t,q)\in D}\min_{1\leq j\leq n}  d_j^n(P\T f(t,q)P )>0\\
    b\coloneqq& \sup_{(t,q)\in D}\max_{2\leq j\leq n}  u_j^n(P\T f(t,q)P)+\epsilon<+\infty, \label{eq:bj}
  \end{align}
  then it follows from \emph{i)} in \Cref{thm:matrix} and the definition \eqref{eq:An} of~$\An(a,b)$ that
  \begin{equation}
    v\T P\T f(t,q)  \leq -\lambda v\T P\T \text{ for all } (t,q)\in D,
  \end{equation}
  where $(\lambda,v)$ is the solution to \eqref{eq:evp}.
\end{remark}
To apply \Cref{rem:matrix}, it is required to verify \eqref{eq:aj}--\eqref{eq:bj} for some permutation matrix $P\in \Pn$.
Since $\Pn$ contains $n!$ elements, this quickly becomes impractical as $n$ increases.
In the next section we study a special case of \eqref{eq:C} for which $P$ can be found by sorting the elements of a vector in $\rn$.

\subsection{Robustness and a Converse Theorem}
\label{sec:converse}
In this section we study a special case of \eqref{eq:C} where $f$ can be written as the sum of a compartmental matrix and a perturbation $p:D\to \Cn$,
\begin{equation}
  \label{eq:Fp}
  f(t,q)=F+p(t,q) \text{ for all } (t,q)\in D.
\end{equation}
Our first goal is to derive an explicit construction of $(\lambda,v)$ such that
\begin{equation}
  \label{eq:Fp-lyap}
  v\T\big(F+p(t,q)\big)\leq -\lambda v\T \text{ for all } (t,q)\in D
\end{equation}
when $F$ is invertible and the off-diagonal elements of $p$ are bounded.
This yields a practical criterion for verifying exponential stability of the null solution to \eqref{eq:C} and shows that the stability condition is robust to bounded perturbations.
Finally, under additional restrictions on $p$,  we establish a converse result showing that \eqref{eq:Fp-lyap} is equivalent to attractivity of the null solution to \eqref{eq:C}.

\begin{assumption}\label{A2}
  \Cref{A1} and \eqref{eq:Fp} hold for a given matrix $F\in \Cn$ and perturbation $p:D\to \Cn$ such that
  \begin{equation}
    \label{eq:p-bounded}
    (F+p(t,q))_{ij}\leq c \text{ for all } (t,q)\in D \text{ and } i\neq j,
  \end{equation}
  for a fixed constant $c>0$.
\end{assumption}
The following proposition provides an algorithmic procedure to construct $\lambda>0$ and $v\in(0,+\infty)^n$ so that \eqref{eq:Fp-lyap} holds.
\begin{proposition}\label{prop:formula}
  Suppose \Cref{A2} holds and that $F$ is invertible.
  Define $z\T \coloneqq - \bm 1\T F^{-1}$ and define $P\in \Pn$ such that
  \begin{equation}
    \label{eq:zorder}
    (P\T z)_1\geq \dots \geq (P\T z)_n.
  \end{equation}
  Let $(\lambda,w)$ be the solution to \eqref{eq:evp} with $a$, $b$ given by
  \begin{equation}
    \label{eq:ab-def}
    a\coloneqq \min_id_i^n(P\T FP),\quad b\coloneqq (n-1)c.
  \end{equation}
  Then inequality \eqref{eq:Fp-lyap} holds with $v\coloneqq Pw$.
\end{proposition}
\begin{remark}
  Since $z\in \rn$, there always exists a permutation matrix $P$ that sorts the elements of $z$ in accordance with \eqref{eq:zorder}.
\end{remark}

Before we state the converse result, we introduce some notation and definitions.
\begin{notation}
  For $F\in \Cn$, let $F_{0j}\coloneqq -(\bm 1\T F)_j$, $j=1,\dots,n$.
\end{notation}
\begin{definition}
  For $F\in \Cn$, define the index sets
  \begin{align}
    \mathcal I_n&\coloneqq \{0,\dots,n\}\times \{1,\dots,n\},\\
    \mathcal Z_n(F)&\coloneqq \{(i,j)\in \mathcal I_n : F_{ij}= 0,\, i\neq j\}.
  \end{align}
\end{definition}
In the following theorem, we restrict the perturbation to satisfy
\begin{equation}\label{eq:ZpF}
  \mathcal Z_n(F)\subset \mathcal Z_n(p(t,q)),\quad (t,q)\in D.
\end{equation}
This means that $p$ does not perturb those off-diagonal elements and column sums of $F$ that are equal to zero.
\begin{theorem}\label{thm:converse}
  Consider \eqref{eq:C} with \Cref{A2} and suppose \eqref{eq:ZpF} holds.
  Then the following statements are equivalent:
  \begin{enumerate}
 \item the null solution to \eqref{eq:C} is attractive with respect to $\X$;
 \item the matrix $F$ is nonsingular;
 \item there are $\lambda>0$ and $v\in(0,+\infty)^n$ such that inequality \eqref{eq:Fp-lyap} holds; and
 \item the null solution to \eqref{eq:C} is ES with respect to $\X$.
  \end{enumerate}
\end{theorem}

\section{Numerical Example}\label{sec:sim}
To illustrate the theory presented in \Cref{sec:main} we consider the system
\begin{equation}
  \label{eq:E}\tag{E}
  \dot x(t) = f(t,x(t))x(t),
\end{equation}
where $f:\rb\times \rb^3\to\rb^{3\times 3}$ is given by
\begin{equation}\label{eq:f}
  f(t,q)\coloneqq
 \begin{bmatrix}
   -\frac{1}{1+|q|_1} & |q|_1 &0\\
0 & -\frac{1}{1+|q|_1}-|q|_1 &|q|_1\\
      \frac{1}{1+|q|_1} & \frac{1}{1+|q|_1}  &-\frac{1}{1+|q|_1}-|q|_1
\end{bmatrix}.
\end{equation}
\Cref{A1} is satisfied for the system \eqref{eq:E} with $\tilde D\coloneqq \rb\times \rb^3$ and $\X\coloneqq [0,+\infty)^3$.
It follows that for each $s\geq 0$, the set
\begin{equation}
  \label{eq:S(s)}
  \Set(s)\coloneqq \{q\in [0,+\infty)^3\;:\; |q|_1\leq s\}\subset \rb^3
\end{equation}
is forward invariant for \eqref{eq:E}, since $\sum_{i=1}^3(f(t,q)q)_i\leq 0$ for all $(t,q)\in \tilde D$.
In what follows, we fix $s>0$ and assess the stability of the null solution to \eqref{eq:E} in view of \Cref{thm:matrix} with respect to $\Set(s)$.

Recall the definitions \eqref{eq:dj} and \eqref{eq:uj} of $d_j^n$ and $u_j^n$, and the hypotheses of \Cref{thm:matrix}.
Straightforward calculations show that
  \begin{alignat}{2}
    d_j^3(f(t,q))&=\frac{1}{1+|q|_1}\geq\frac{1}{1+s},\quad &&j=1,2,3,\\
    u_j^3(f(t,q))&=|q|_1\leq s,\quad &&j=2,3,
  \end{alignat}
for all $t\geq 0$ and $q\in \Set(s)$.
Therefore,
\begin{equation}
  \label{eq:35}
  f(t,q)\in \A_3\big(a,b),\quad q\in \Set(s)
\end{equation}
with $a\coloneqq \frac{1}{1+s}>0$, $b\coloneqq s$, and $\A_3(a,b)$ as defined in \eqref{eq:An}.
Since $\Set(s)$ is forward invariant, we conclude by \Cref{thm:matrix} and \Cref{prop:lyap} that the null solution to \eqref{eq:E} is ES with respect to $\Set(s)$.
An exponential decay rate $\lambda$ and the vector $v$ are obtained from the eigenvalue problem \eqref{eq:evp}.\footnote{The eigenvalue problem \eqref{eq:evp} is well-posed by \Cref{lem:gn}.}
Since the parameter $s>0$ can be chosen arbitrarily large, the null solution to \eqref{eq:E} is attractive with respect to $[0,+\infty)^3$.

\begin{figure}[h!]
    \centering
    \includegraphics[width=0.5\textwidth]{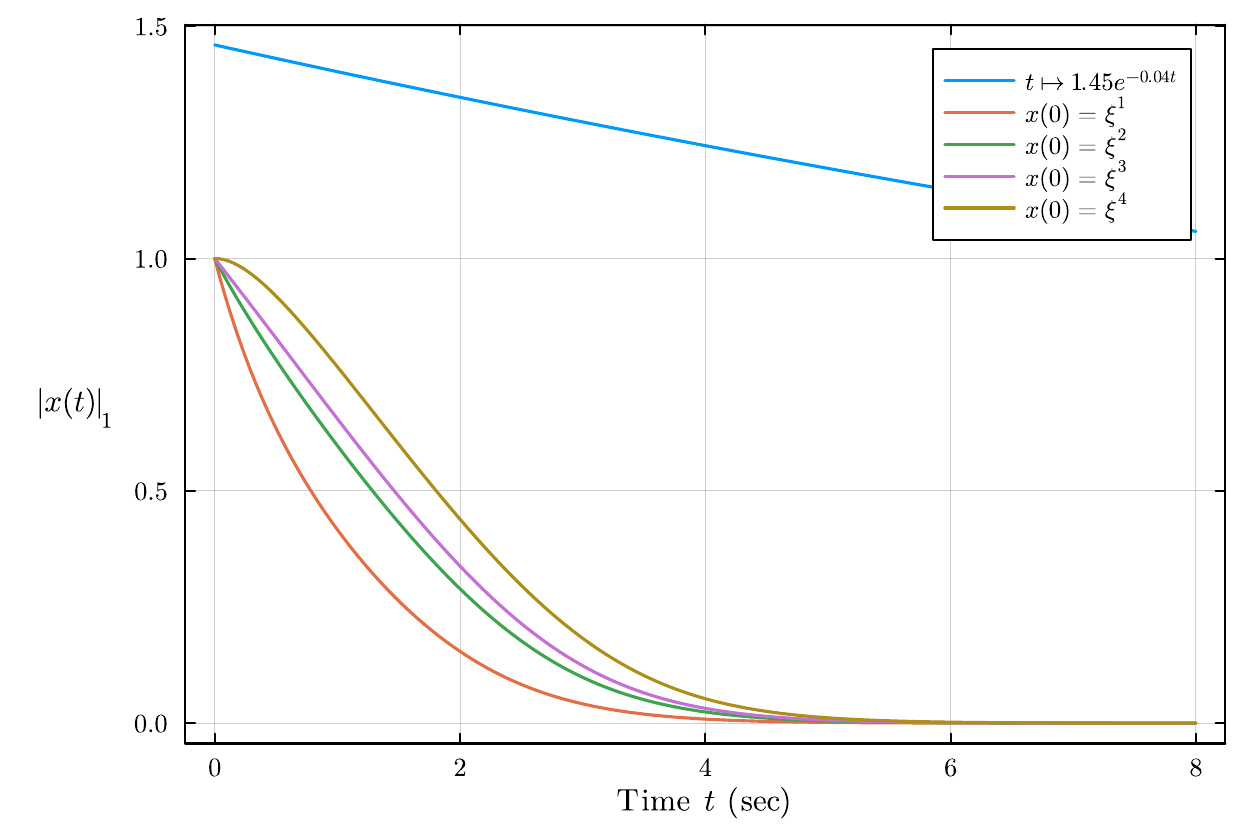}
    \caption{Time evolution of $|x(t)|_1$ for solutions $x$ to \eqref{eq:E} with the initial conditions $\xi^1=(0,0,1)$, $\xi^2=(0,1/2,1/2)$, $\xi^3=(1/3,1/3,1/3)$, and $\xi^4=(1,0,0)$ in $\Set(s)$.
      The overbound \eqref{eq:64} is satisfied for each initial condition.}
    \label{fig:plt}
  \end{figure}

Setting $s=1$ yields the parameters $a=\frac{1}{2}$ and $b=1$.
By solving \eqref{eq:evp} numerically we obtain
\begin{equation}
  \label{eq:21}
  \lambda = 0.04,\quad v=(1.0,0.92,0.69)
\end{equation}
with the corresponding ratio $\frac{\max_{i} v_i}{\min_{i} v_i}= 1.45$.  
\Cref{prop:lyap} and \Cref{thm:matrix} ensure that every solution $x$ to \eqref{eq:E} starting at~$\xi$~in~$\Set(1)$ satisfies
\begin{equation}
  \label{eq:64}
  |x(t)|_1\leq 1.45 e^{-0.04 t}|\xi|_1,\quad t\geq 0,
\end{equation}
which is numerically verified in \Cref{fig:plt} for a selection of initial conditions.
Observe that the exponential decay rate $\lambda =0.04$ is conservative over the time interval $[0,8]$.
This concludes the numerical example.
  
\section{Conclusion}\label{sec:conclusion}
This technical note considers exponential stability of the null solution for a class of compartmental systems using linear Lyapunov functions.
Sufficient conditions under which such functions exist are provided, with a focus on constructive procedures to obtain them.
In the general case, the construction of the Lyapunov functions involves finding a permutation matrix by brute force.
In a special case, the permutation matrix can be found by sorting the elements of a vector in $\rn$, rather than by brute force.
For this special case, an equivalence between exponential stability and invertibility of an associated matrix is established; the invertibility condition can be verified numerically with little effort.

\appendix
\subsection{Notation}
\begin{table}[H]
          \caption{Notation summary used throughout the paper.}
        \begin{tabular*}{\linewidth}{@{\extracolsep{\fill}} l c }
          Symbol & Meaning or Definition\\
          $\Pn$ & The set of $(n\times n)$ permutation matrices\\
          $\Cn$ & The set of $(n\times n)$ compartmental matrices \\
          $\Dn$ & The set of downstream connected $F\in \Cn$\\
          $\An(a,b)$ & Specific subset of $\Dn$, see \Cref{thm:matrix}\\
          $f=(f_{ij})$ & Matrix function (cf. \eqref{eq:C})\\
          $\tilde D$ & Subset of $\rb^{n+1}$, domain of $f$\\
          $D$ & Subset of $\tilde D$, defined as $D\coloneqq [0,+\infty)\times \X$\\
          $f_{0j}$ & $j$-th negated column sum of $f$\\
          $g_n$ & $(n\times n)$ matrix function, defined in \eqref{eq:gn}\\
          $\X$ & Subset of $[0,+\infty)^n$, forward invariant for \eqref{eq:C}\\
          $x$ & State vector of \eqref{eq:C}\\
          $t$ & Independent variable time\\
          $d^n_j$, $u^n_j$ & Functions defined in \eqref{eq:dj}, \eqref{eq:uj}\\
          $q$ & Vector in $\rn$, typically in $\X$\\
          $v$ & Vector in $(0,+\infty)^n$\\
          $\lambda$ & Positive real number, exponential decay rate\\
          $|\cdot|_1$ & $1$-norm\\
        \end{tabular*}
\end{table}

\subsection{Supporting lemmas}\label{appendix:A}
Before we state the lemmas we define some terms about matrices.
Let $F$ be a matrix in $\rnn$.
Then $F$ is said to be \emph{Hurwitz} if the real part of every eigenvalue of $F$ is negative.
If every off-diagonal element of $F$ is nonnegative, then $F$ is called \emph{Metzler} \cite{deLeenheer2001}.
The matrix $F$ is said to be \emph{reducible} \cite[Definition 6.2.21]{horn2012} if there exists a permutation matrix $P\in \rnn$ such that $P\T FP$ is block triangular; otherwise, $F$ is said to be \emph{irreducible}.
\begin{lemma}\label{lem:eigen}
  If $F\in \rnn$ is nonsingular and compartmental, then $F$ is Hurwitz.
\end{lemma}

\begin{proof}
  See Theorem 3 and item (a) at page 52 in \cite{jacquez93}.
\end{proof}
The following lemma provides the formula for $\lambda$, $v$ mentioned in \Cref{rem:lyap} and is used in the proof of \Cref{prop:formula}.
\begin{lemma}[\cite{fife1972}, \cite{rantzer2015}]\label{lem:support}
  If $w\in (0,+\infty)^n$ and  $F\in \rnn$ is both compartmental and nonsingular, then
  \begin{equation}\label{eq:v-lambda}
    v\T \coloneqq -w\T F^{-1}\in(0,+\infty)^n,\quad v\T F\leq -\lambda v\T,
  \end{equation}
  where $\lambda \coloneqq \min_{i}\frac{w_i}{v_i}>0$.
\end{lemma}
\begin{proof}
  Since $F$ is compartmental and nonsingular, $F$ is Hurwitz by \Cref{lem:eigen}.
  Note that a compartmental matrix is also Metzler.
  Now it follows from Proposition 1 in \cite{rantzer2015} that $-F^{-1}$ exists and has only nonnegative elements.
  In particular, every column of $-F^{-1}$ has at least one positive element.
  Therefore the vector $v$ in \eqref{eq:v-lambda} has only positive elements.
  Hence,
  \begin{equation}
    v\T F= -w\T F^{-1} F=-w\T\leq -\lambda v\T,
  \end{equation}
  where the last step follows by definition of $\lambda$.
\end{proof}
The next lemma is used to prove \Cref{thm:matrix} and the first statement $1)$ is due to the Perron--Frobenius Theorem.
\begin{lemma}\label{lem:gn}
  Fix any integer $n>1$ and real numbers $a,b>0$ and let $g_n(a,b)\in \rnn$ be given by \eqref{eq:gn}.
  \begin{enumerate}
  \item The problem \eqref{eq:evp} has a unique solution.
  \item If $(\lambda,v)$ solves \eqref{eq:evp}, then $v_1>\dots >v_n>0$.
  \end{enumerate}
\end{lemma}
\begin{proof}
  We first prove statement 1).
  Denote $F\coloneqq g_n(a,b)$.
  First we prove that there exists a solution to \eqref{eq:evp}.
  Let $\Gamma$ denote the directed graph of $F$.
  Since $\Gamma$ contains the directed arcs $(1,n)$ and $(i+1,i)$ for $i=1,\dots,n-1$, it follows that $\Gamma$ is strongly connected \cite[Definition 6.2.13]{horn2012}.
  Hence $F$ is irreducible by Theorem 6.2.24 in \cite{horn2012}.
  Since $F$ is compartmental, irreducible, and the column sum $\sum_{i=1}^nF_{in}=-a$ is negative, it follows from \Cref{lem:eigen} and Proposition 1 in \cite{deLeenheer2001} that $F$ is Hurwitz.

  Let $\alpha>0$ be large enough that the elements of $G\coloneqq F+\alpha I$ are nonnegative.
  Since $F$ is irreducible, $G$ is also irreducible.
  Then it follows from \cite[Theorem 1.4 p. 27]{berman1994} that there exists $z\in(0,+\infty)^n$ such that
  \begin{equation}
    \label{eq:perron}
   z\T G=z\T(F+\alpha I)=\rho(G) z\T
  \end{equation}
  where $\rho(G)\geq 0$ is the spectral radius of $G$, $\rho(G)$ is a simple eigenvalue of $G$, and every left eigenvector of $G$ in $[0,+\infty)^n$ is a scalar multiple of $z$.

  By rewriting \eqref{eq:perron} we obtain
  \begin{equation}
    \label{eq:10}
    z\T F = -(\alpha-\rho(G))z\T
  \end{equation}
  where $\lambda \coloneqq (\alpha-\rho(G))$ must be positive, since $F$ is Hurwitz.
  It now follows that $(\lambda,v)$ with $v\coloneqq z/z_1$ is a solution to \eqref{eq:evp}.
  
  To prove uniqueness, fix any two solutions $(\lambda,v)$ and $(\sigma,w)$ to \eqref{eq:evp} and write
  \begin{equation}
    \label{eq:13}
    \begin{split}
      v\T G&=v\T (F+\alpha I)=(\alpha-\lambda) v\T,\\
      w\T G&=w\T (F+\alpha I)=(\alpha-\sigma) w\T.
    \end{split}
  \end{equation}
  Since $w,v\in(0,+\infty)^n$ it follows that $w$, $v$  are scalar multiples of $z$.
  Hence $\alpha-\lambda = \rho(G)=\alpha-\sigma$, which implies that $\lambda = \sigma$.
  Finally, since $\rho(G)$ is a simple eigenvalue of $G$ and $v_1=w_1$ it follows that $w=v$ and this proves statement 1).

  Now we prove 2).
  Fix any $\lambda>0$, $v\in(0,+\infty)^n$ such that $v\T g_n(a,b)=-\lambda v\T $.
  We need to show that $v_1>\dots>v_n$.

  Using the definition \eqref{eq:gn} of $F$, we can write
  \begin{align} \label{eq:vF11}
    (v\T F)_1&=-\lambda v_1= -av_1+av_2<0,\\\label{eq:vFii}
    (v\T F)_i&=-\lambda v_i= bv_1-(a+b)v_i+av_{i+1}<0
  \end{align}
  for $i=2,\dots,n-1$.
  Since $a>0$, it follows from \eqref{eq:vF11} that $v_1>v_2$.
  The inequality \eqref{eq:vFii} can be rewritten as
\begin{equation}
    \label{eq:11}
    v_i-v_{i+1}>\frac{b}{a}(v_1-v_i),\quad i=2,\dots,n-1.
  \end{equation}
  Fix any $i\in \{2,\dots,n-1\}$ and suppose  $v_1>v_2>\dots>v_i$.
  Then it follows from \eqref{eq:11} that $v_1>\dots>v_i>v_{i+1}$.
  Since $v_1>v_2$, it follows by induction that $v_1>v_2>\dots >v_n$.
  This proves statement 2) and completes the proof of the lemma.
 \end{proof}

\subsection{Proof of the Main Results}
 \subsubsection{Proof of \Cref{prop:lyap}}
 Fix any initial condition $\xi\in \X$ and let $x$ be the solution to \eqref{eq:C} such that $x(0)=\xi$.
 The proposition follows if
 \begin{equation}
   \label{eq:12}
   |x(t)|_1\leq \frac{\max_i v_i }{\min_i v_i } e^{-\lambda t}|x(0)|_1,\quad t\geq 0.
 \end{equation}
 It follows from \Cref{A1} that $x(t)\in \X\subset [0,+\infty)^n$ for all $t\geq 0$.
 Define the function $t\mapsto V(t)\coloneqq v\T x(t)$, $t\geq 0$.
 By \eqref{eq:lyap}, it follows that $\dot V(t)=v\T f(t,x(t))x(t)\leq -\lambda V(t)$, $t\geq 0$ since $x(t)\in \X\subset[0,+\infty)^n$, $t\geq 0$.
 It now follows from Gr\"onwall's inequality that
 \begin{equation}
   \label{eq:9}
    V(t)\leq e^{-\lambda t}V(0), \quad t\geq 0.
 \end{equation}
 Since $x(t)\in [0,+\infty)^n$ for all $t\geq 0$, we can write $\sum_{i=1}^nx_i(t)=|x(t)|_1$ for all $t\geq 0$.
 By the latter, the definition of $V$, and \eqref{eq:9} we obtain
 \begin{equation}
   \begin{split}
     V(0) &\leq (\max_i v_i ) |x(0)|_1\\
     |x(t)|_1&\leq \frac{1}{\min_i v_i}V(t)\leq \frac{1}{\min_i v_i}e^{-\lambda t}V(0) ,\quad t\geq 0.
   \end{split}
 \end{equation}
 since $v\in(0,+\infty)^n$, which in turn implies \eqref{eq:12}.
 This concludes the proof.
 \qed
 \subsubsection{Proof of \Cref{prop:ds}}
 We can without loss of generality assume that $P$ is the identity matrix $I$ and that $v_1\geq \dots \geq v_n$.
 Note that $P \F P\T\subset \Cn$ for all $P\in \Pn$.
 If $P\neq I$, $(P\T v)_1\geq \dots \geq (P\T v)_n$, we can do the relabeling $\tilde\F \coloneqq P \F P\T$ and $\tilde v \coloneqq P \T v$.
 Since $(\lambda,v)$ solves \eqref{eq:p1} for $\F$ it follows that $(\lambda,\tilde v)$ solves \eqref{eq:p1} for $\tilde \F$.
 This corresponds with the special case $P=I$ and $v_1\geq \dots \geq v_n$.

 Recall \Cref{def:ds} and the definition \eqref{eq:dj} of $d^n$.
 If \eqref{eq:F-permute} holds, then it follows that $\F\subset\Dn$, since $\lambda>0$.
 From now, let $F\in \F\subset \Cn$ be arbitrary but fixed and set $d\coloneqq d^n(F)$ and $u_j\coloneqq u_j^n(F)$, $j=2,\dots,n$.
 It remains to show that
 \begin{equation}\label{eq:dj-F}
   d_j\geq \lambda ,\quad j=1,\dots,n.
 \end{equation}
 It is assumed that $(\lambda,v)$ solves \eqref{eq:p1}.
 If we can show that
 \begin{equation}\label{eq:vFj}
   (v\T F)_j\geq -v_jd_j,\quad j=1,\dots,n
 \end{equation}
 then \eqref{eq:dj-F} follows, since $\lambda>0$ and $(v\T F)_j\leq -\lambda v_j$ and $v_j>0$ for $j=1,\dots,n$.

 By definition \eqref{eq:dj}, \eqref{eq:uj} of  $d_j$ and $u_j$ we can write
 \begin{equation}\label{eq:Fjj}
   F_{11}=-d_1,\quad F_{jj}=-d_j-u_j,\quad j=2,\dots,n.
 \end{equation}
 Recall that $F$ is compartmental, so its column sums are nonpositive and its off-diagonal elements are nonnegative.
 Using $v_1\geq \dots \geq v_n>0$ and the definition of $u_j$ we obtain the bounds
 \begin{align}
   \label{eq:vFu}
   \sum_{i=1}^{j-1}v_iF_{ij}&\geq v_j\sum_{i=1}^{j-1}F_{ij}=v_ju_j,\quad j>1,\\
   \sum_{i=j+1}^nv_iF_{ij}&\geq 0,\quad j<n. \label{eq:vFd}
 \end{align}

 We prove \eqref{eq:vFj} case by case.
 For the case $j=1$ we obtain
 \begin{equation}
     (v\T F)_1=v_1F_{11}+\sum_{i=2}^nv_iF_{i1}\geq -v_1d_1
 \end{equation}
 by using \eqref{eq:vFd} and inserting $F_{11}=-d_1$ in the first step.

 Next we prove the case $j=n$.
 Inserting $F_{nn}$ in \eqref{eq:Fjj} into the definition of $(v\T F)_n$ we get
  \begin{equation}\label{eq:case-n}
     (v\T F)_n= \sum_{i=1}^{n-1}v_iF_{in}-v_nu_n-v_nd_n.
   \end{equation}
 The bound \eqref{eq:vFj} now follows from \eqref{eq:vFu}.

 For the last case, fix any $j\in \{2,\dots,n-1\}$.
 First we write 
 \begin{equation}
   \begin{split}
     (v\T F)_j=&\sum_{i=1}^{j-1}v_iF_{ij}-v_j(u_j+d_j)+\sum_{i=j+1}^nv_iF_{ij}\\
     \geq & \sum_{i=1}^{j-1}v_iF_{ij}-v_j(u_j+d_j)\\
     \geq & v_ju_j-v_j(u_j+d_j)=-v_jd_j
   \end{split}
 \end{equation}
 where we insert $F_{jj}$ from \eqref{eq:Fjj} into the definition of $(v\T F)_j$ in the first step.
 In the second and third step we use \eqref{eq:vFd} and \eqref{eq:vFu} respectively.
 This shows the last case and concludes the proof.
 \qed

 \subsubsection{Proof of \Cref{thm:matrix}}
   We prove \emph{ii)} first, under the assumption that \emph{i)} is true.
   Let $\F\subset \Cn$ be bounded, and suppose \eqref{eq:p1} admits a solution $(\lambda,v)$.
  It then follows from \Cref{prop:ds} that there is $P\in \Pn$ such that $d^n(A)\geq (\lambda,\dots,\lambda)$ for all $A\in P \F P\T$.
  Fix any such $P$ and recall the definition \eqref{eq:uj} of the map $u_j^n$.
  Since $\F$ is bounded, there is $b>0$ such that $u_j^n(A)\leq b$ for all $j=2,\dots,n$ and $A\in P\T \F P$.
  It now follows from \emph{i)} that $P\T \F P\subset \An(\lambda,b)$.

  Suppose conversely that $P \F P\T\subset \An(a,b)$ for some fixed $a,b>0$.
  Let $(\lambda,v)$ be the solution to \eqref{eq:evp}.
  Then it follows by definition \eqref{eq:An} of $\An(a,b)$ that $(\lambda,P\T v)$ solves \eqref{eq:p1} for $\F$.

  Now we prove \emph{i)}.
  Fix any $a,b>0$ and $F\in \Cn$ such that \eqref{eq:djuj} holds.
  Let $(\lambda,v)$ be the solution to \eqref{eq:evp}, which is well-defined by \Cref{lem:gn} and satisfies $v_1>\dots>v_n>0$.

  It follows by the definition \eqref{eq:gn} of $g\coloneqq g_n(a,b)$ that we can write
  \begin{equation}
    \label{eq:vG}
    \begin{split}
      (v\T g)_1&=-a(v_1-v_2),\\
      (v\T g)_j&= b(v_1-v_j)-a(v_j-v_{j+1}),\quad j\neq 1,n\\
      (v\T g)_n&= b(v_1-v_n)-av_n.
    \end{split}
  \end{equation}
  And it follows by the definition \eqref{eq:An} of $\An(a,b)$, that $F\in \An(a,b)$ if
  \begin{equation}
    \label{eq:vFG}
    (v\T F)_j\leq (v\T g)_j,\quad j=1,\dots,n.
  \end{equation}
  since $v\T g=-\lambda v\T$.

  We divide the proof of \eqref{eq:vFG} into the cases $j=1$, $j=n$, and $j\in\{2,\dots,n-1\}$.
  First we derive some inequalities that are used in all the cases.
  Recall the definitions \eqref{eq:dj} and \eqref{eq:uj} of $d^n$ and $u_j^n$.
  Since $F$ is fixed, we set $d\coloneqq d^n(F)$ and $u_j=u_j^n(F)$, $j=2,\dots,n$.
  The inequality
  \begin{equation}
    \label{eq:uj-bound}
    \sum_{i=1}^{j-1}v_iF_{ij}\leq v_1\sum_{i=1}^{j-1}F_{ij}=v_1u_j,\quad j>1
  \end{equation}
  follows from the ordering $v_1>\dots>v_n$ of $v$ and because $F_{ij}\geq 0$ for $i\neq j$ since $F\in \Cn$, and the definition of the map $u_j^n$.
  The inequality
\begin{equation}
  \label{eq:dj-bound}
    \sum_{i=j+1}^nv_iF_{ij}\leq v_{j+1}\sum_{i=j+1}^nF_{ij} \leq  v_{j+1}d_j^{n},\quad j<n
\end{equation}
follows from the ordering of $v$, because $F\in \Cn$, and the definition of the map $d_j^n$.

Now we prove the case $j=1$ of \eqref{eq:vFG}.
We obtain
\begin{equation}
  \begin{split}
    (v\T F)_1=& v_1 F_{11}+\sum_{i=2}^nv_iF_{i1}\\
    \leq& -d_1v_1+v_2d_1=-d_1(v_1-v_2)
  \end{split}
\end{equation}
where we use $F_{11}=-d_1$ and \eqref{eq:dj-bound} in the second step.
Since $v_1>v_2$ and $d_1\geq a$ by assumption, we obtain $(v\T F)_1\leq -a(v_1-v_2)=(v\T g)_1$.

For the second case, fix any $j\in\{2,\dots,n-1\}$.
We compute
\begin{equation}
  \begin{split}
    (v\T F)_j=&\sum_{i=1}^{j-1}v_{i}F_{ij}+v_jF_{jj}+\sum_{i=j+1}^nv_i F_{ij}\\
    \leq & v_1u_j-v_j(u_j+d_j)+v_{j+1}d_j\\
    = &u_j(v_1-v_j)-d_j(v_j-v_{j+1})\\
    \leq & b (v_1-v_j)-a(v_j-v_{j+1})
  \end{split}
\end{equation}
using \eqref{eq:dj-bound}, \eqref{eq:uj-bound} and the formula $F_{jj}=-d_j-u_j$ in the second step.
The last step follows from the ordering of $v$ and the assumption \eqref{eq:djuj}.
This shows \eqref{eq:vFG} by inspection of \eqref{eq:vG}.

Lastly we prove the case $j=n$.
Using similar steps as in the second case we obtain
\begin{equation}
  \label{eq:1}
  (v\T F)_n\leq u_n(v_1-v_n)-d_nv_n\leq b(v_1-v_n)-av_n=(v\T g)_n
\end{equation}
where the second step follows from \eqref{eq:djuj} and since $v_1>v_n>0$.
We have now shown \eqref{eq:vFG} for all $j\in\{1,\dots,n\}$.
This proves $i)$ and concludes the proof of the theorem.
\qed

\subsubsection{Proof of \Cref{prop:formula}}
If we can show that
\begin{equation}
  \label{eq:20}
  \tilde f(t,q)\coloneqq P\T (F+p(t,q))P\in \An(a,b),\quad (t,q)\in D,
\end{equation}
then the proposition follows by the definition \eqref{eq:An} of $\An(a,b)$.

Since $F$ is nonsingular and compartmental, it follows from \Cref{lem:support} that $z\in (0,+\infty)^n$ and $z\T F\leq -\tilde \lambda z\T$ for some $\tilde \lambda>0$.
Since $(P\T z)_1\geq \dots \geq (P\T z)_n$, it follows from \Cref{prop:ds} applied to $\F\coloneqq \{F\}$ that
\begin{equation}
  \label{eq:18}
  d^n(P\T F P)\geq (a,\dots,a),
\end{equation}
by the definition \eqref{eq:ab-def} of $a$, where $d^n$ is defined in \eqref{eq:dj}.
By definition of $d^n$ and $\tilde f$, and inequality \eqref{eq:18}, we obtain
\begin{equation}
  \label{eq:19}
  d^n(\tilde f(t,q))\geq (a,\dots,a),\quad (t,q)\in D,
\end{equation}
since $P\T p(t,q)P\in \Cn$ for all $(t,q)\in D$ as $P$ is a permutation matrix.
Using the bound \eqref{eq:p-bounded} in \Cref{A2} and the definition of $\tilde f$, it follows that
\begin{equation}
  \label{eq:16}
  u^n_j(\tilde f(t,q))\leq (n-1)c=b,\quad (t,q)\in D,
\end{equation}
since $P$ is a permutation matrix.
It now follows from \Cref{thm:matrix} that $\tilde f(t,q)\in \An(a,b)$ for all $(t,q)\in D$, which completes the proof.
\qed
\subsubsection{Proof of \Cref{thm:converse}}
The theorem follows once we have shown the chain of implications $1)\implies 2)\implies 3) \implies 4) \implies 1)$.
To show  $1)\implies 2)$, we prove the contrapositive statement: if $F$ is singular, then the null solution to \eqref{eq:C} is not attractive with respect to $\X$.

 Suppose $F$ is singular and let $x$ be the solution to \eqref{eq:C} starting at $x(0)=\xi$ for some $\xi\in(0,+\infty)^n$.
 Since $x(t)\in \X\subset [0,+\infty)^n$ for all $t\geq 0$, cf. \Cref{A1}, it is enough to find $w\in\{0,1\}^n\setminus\{0\}^n$ such that
 \begin{equation}
   \label{eq:4}
   \big(w\T (F+p(t,q))\big)_j\geq 0,\quad (t,q)\in D,\quad j=1,\dots,n.
 \end{equation}
 Indeed, if \eqref{eq:4} holds, then $w\T \dot x(t)=w\T(F+p(t,x(t)))x(t)\geq 0$ for all $t\geq 0$ which implies that $|x(t)|_1\geq w\T x(t)\geq w\T \xi>0$ for all $t\geq 0$.
 Hence, the null solution to \eqref{eq:C} is not attractive with respect to $\X$.
 
 In the construction of $w$ we consider two cases.
 In the first case, assume that every column sum of $F$ is zero.
 Then every column sum of $p(t,q)$ is zero for all $(t,q)\in D$ by the assumption \eqref{eq:ZpF}.
 Therefore, \eqref{eq:4} holds with $w\coloneqq (1,\dots,1)$.

 In the second case, assume that $F$ has at least one negative column sum.
 Since $F$ is compartmental, is singular, and has a negative column sum, it follows from \cite{fife1972} that there exists $P\in \Pn$ such that
 \begin{equation}
   \label{eq:5}
   P F P \T =
   \begin{bmatrix}
     U & 0 \\
     Q & R
   \end{bmatrix}
 \end{equation}
 where $U$ and $R$ are square matrices, $0$ denotes a zero matrix, and the column sums of $R$ are all zero.
 Let $P$ be such a permutation matrix.
 Let $(r\times r)$ denote the dimension of $R$ and define $v\in\{0,1\}^n$ as follows
 \begin{equation}
   \label{eq:2}
   v_i\coloneqq
   \begin{cases}
     0,\quad i=1,\dots,n-r,\\
     1,\quad \text{else}.
   \end{cases}
 \end{equation}
 Then $v\T P F P\T\in[0,+\infty)^n$ since every column sum of $R$ is zero and $Q$ is nonnegative, as the elements of $Q$ correspond to off-diagonal elements of $F$.
 Setting $w\T \coloneqq v\T P$, we have that $w\T F$ is a vector with nonnegative elements, $w\in\{0,1\}^n$, and $w\neq (0,\dots,0)$.

 Now that $w$ is fixed, fix any $(t,q)\in D$ and write, with slight abuse of notation, $p\coloneqq p(t,q)$.
 If we can show that $w\T p\geq 0$, then \eqref{eq:4} follows and we are done.
 Define $p_{0j}\coloneqq -\sum_{i=1}^np_{ij}$, $F_{0j}\coloneqq -\sum_{i=1}^nF_{ij}$, for $j=1,\dots,n$.
 Then we can write 
 \begin{equation}
   \label{eq:14}
   \begin{split}
     (w\T p)_j=&-w_jp_{0j}-\sum_{i\neq j}(w_j-w_i)p_{ij},\quad \\
     (w\T F)_j=&-w_jF_{0j}-\sum_{i\neq j}(w_j-w_i)F_{ij},\quad 
   \end{split}
 \end{equation}
 for $j=1,\dots,n$.
 
 Suppose for the sake of contradiction, that there exists $j\in \{1,\dots,n\}$ such that $(w\T p)_j<0$ and keep in mind that $(w\T F)_j\geq 0$.
 Remember that $F$ and $p$ are compartmental matrices, so $F_{ij},p_{ij}\geq 0$ for $i\neq j$.
 Therefore, if $(w\T p)_j<0$, it must be that $w_j=1$.
 If $w_j=1$, it follows by inspection of \eqref{eq:14} that $F_{0j}$ must be zero since $(w\T F)_j\geq 0$.
 This implies in turn that $p_{0j}=0$ by the assumption \eqref{eq:ZpF}.
 For $(w\T p)_j$ to be negative, it must be that $w_j=1$, $p_{0j}=0$, and $F_{0j}=0$.
 We can thus write
 \begin{equation}
   \label{eq:17}
   \begin{split}
     (w\T p)_j=&-\sum_{i\neq j}(w_j-w_i)p_{ij},\\
     (w\T F)_j=&-\sum_{i\neq j}(w_j-w_i)F_{ij}.
   \end{split}
 \end{equation}
 Since $(w\T p)_j<0$, there is $k\neq j$ such that $w_k=0$ and $p_{kj}>0$.
 Since $p_{kj}>0$ it follows that $F_{kj}>0$, by the assumption \eqref{eq:ZpF}.
 This implies in turn that $(w\T F)_j<0$, by inspection of \eqref{eq:17}, contradicting the fact that $(w\T F)_j\geq 0$.
 This concludes the proof of $1)\implies 2)$.

 The implication $2)\implies 3)$ follows from \Cref{prop:formula}.
 The remaining implications $3)\implies 4)$ and $4)\implies 1)$ follow from \Cref{prop:lyap} and \Cref{def:stability} respectively.
 This concludes the proof.
 \qed

\bibliographystyle{IEEEtran}
\bibliography{refs}


\end{document}